\documentclass[amsmath,amssymb,twocolumn,prx,superscriptaddress]{revtex4-1}
\usepackage{graphicx}
\usepackage{dcolumn}
\usepackage{subfigure}
\usepackage{bm}
\usepackage{color}
\usepackage{braket}
\usepackage{hyperref}

\begin{document}

\title{Controlling the configuration space topology of mechanisms}
\author{M. Berry}
\affiliation{Department of Physics, Syracuse University, Syracuse, NY,  13244 USA}
\author{David Limberg}
\affiliation{Department of Polymer Science and Engineering, University of Massachusetts, Amherst, MA 01003 USA}
\author{M.E. Lee-Trimble}
\affiliation{Department of Physics, University of Massachusetts, Amherst, MA 01003 USA}
\author{Ryan Hayward}
\affiliation{Department of Chemical and Biological Engineering, University of Colorado, Boulder, CO 80309 USA}
\author{C.D. Santangelo}
\email{cdsantan@syr.edu}
\affiliation{Department of Physics, Syracuse University, Syracuse, NY,  13244 USA}

\date{\today}

\begin{abstract}
Linkages are mechanical devices constructed from rigid bars and freely rotating joints studied both for their utility in engineering and as mathematical idealizations in a number of physical systems. Recently, there has been a resurgence of interest in designing linkages to perform certain tasks from the physics community. We describe a method to design the topology of the configuration space of a linkage by first identifying the manifold of critical points, then perturbing around such critical configurations. We then demonstrate our procedure by designing a mechanism to gate the propagation of a soliton in a Kane-Lubensky chain of interconnected rotors.
\end{abstract}

\maketitle

Linkages serve as prototypical mechanical models for many different physical systems, including animal limbs and joints \cite{burgess2021review,  roos2009linking,  liu2017kinematic}, polymer physics \cite{mazars1996statistical}, protein allostery \cite{sljoka2012algorithms, sljoka2011predicting, hespenheide2004structural, jacobs2001protein, thomas2007simulating}, DNA rigidity \cite{marras2015programmable, lei2018three} origami \cite{liu2018topological,tachi2009simulation,liu2017nonlinear} and jamming \cite{sirono2011dilatancy, connelly2014ball, damavandi2022energetic}. At a basic level, a linkage is a graph whose edges have a fixed length but whose vertices are otherwise freely rotating joints. Yet this superficial simplicity belies behavior that can be surprisingly complex. One of the first important mathematical results was  Kempe's universality theorem, which showed that a linkage can be designed such that a given vertex traces out a portion of any rational algebraic curve. \cite{kempe1875general, li2018kempe}. 

The results of following the proof's design procedures can be unwieldy for even simple curves, yet there are many applications where the precise motion of the vertices of a linkage is less important than the motion's qualitative features. An example is the celebrated Kane-Lubensky (KL) chain \cite{kane2014topological}, a series of rotors joined by springs, which supports the propagation of a soliton called a ``spinner'' in which each rotor, in turn, rotates a full $360^\circ$ degrees \cite{chen2014nonlinear}. The existence and behavior of this soliton is robust under length changes to the rotors, depending on the topology of the configuration space rather than its particular shape \cite{lo2021topology}. Additionally, many linkages have branched configuration spaces, meaning that many different qualitative motions are accessible.  For example, generic origami and kiragami mechanisms have highly branched configuration spaces, leading to pluripotency \cite{hawkes2010programmable, chen2018branches}. Similar branched configuration spaces have been used to design mechanical logic devices \cite{merkle2018mechanical} and kinematotropic mechanisms that can change how many degrees of freedom they can access \cite{wohlhart1996kinematotropic, galletti2001single}. For flexible or imperfectly fabricated mechanisms, in which the fine structure of the motion cannot be controlled anyway, understanding how the topology of their configuration spaces relate to qualitative motions is crucial.

In this paper, we introduce an approach to linkage design that explores this relationship by focusing on the critical points of a configuration space \cite{kumar2000computation, fernandez2012second}. At a critical point, a mechanism has an anomalously large class of potential linear motions available to it, but higher order corrections from the mechanical constraints restrict the motion to a subset of these motions \cite{muller2019singular, connelly1994higher, connelly1996second, muller2018higher}. Critical points are delicate; even small perturbations of the mechanism geometry will destroy them. However, by carefully controlling those perturbations, we show that we can construct a mechanism that allows some control over the configuration space's topology. Our design approach can be summarized in two steps: (1) design a mechanism with a branched configuration space, then (2) perturb the mechanism geometry away from the branched configuration space to a smooth one with controlled topology.
To illustrate this approach, we will apply our design methodology to the KL chain. By replacing one of the unit cells with a designed mechanism, we show that the propagation of the spinner soliton can be controllably gated.

In Sec. \ref{sec:criticalpoints}, we review relevant parts of rigidity theory and mechanisms. In Sec. \ref{sec:controllingtopology}, we describe mathematical tools that provide a geometrical interpretation to critical points. This interpretation will provide the basis of our design methodology, which we will illustrate with an example containing five bars. Finally, in Sec. \ref{sec:gatedchain}, we will use our formalism to explicitly design a mechanism to gate the KL chain. Importantly, the operation of the resulting gate is robust with respect to small perturbations. Finally, we conclude with a brief discussion highlighting new directions enabled by this work.

\section{Critical points in mechanisms}
\label{sec:criticalpoints}

\subsection{Mathematical rigidity}
In this section, we review the basic mathematical description of mechanisms. Though we focus on linkages, which are constructed entirely from free-rotating joints and inextensible bars, the formalism can be generalized to mechanisms with other holonomic constraints. We define a linkage as a collection of $V$ vertices in $d$ dimensions joined by $E$ rigid bars. The configuration of a linkage can then always be represented by a point, $\mathbf{u}$, in the space of vertex positions, which we will denote as $\mathcal{M}$, and has dimension $M = V d$. We assume there are $E$ bars in the linkage and denote the length of the $\alpha^{th}$ bar, $\ell_\alpha(\mathbf{u})$. The configuration space of the linkage can then be represented by the family of equations,
\begin{equation}
\label{eq:constraints}
	\ell^2_\alpha(\mathbf{u}) = L_\alpha^2
\end{equation}
where $L_\alpha$ is the target length of the $\alpha^{th}$ bar. Note that Eq. (\ref{eq:constraints}) is written using the square of $\ell_\alpha(\mathbf{u})$ so that it is can be an analytic function everywhere. By replacing $\ell_\alpha(\mathbf{u})$ with a more general class of functions in Eq. (\ref{eq:constraints}), we can also describe mechanisms with more complex components beyond rigid bars.

Rather than analyzing the configuration space for specific values of $L_\alpha$, we will instead analyze the entire family of configuration spaces that can occur with a fixed network topology by changing the $L_\alpha$. Between Kempe's universality theorem and the potential arbitrariness of $\ell_\alpha(\mathbf{u})$, however, it is indeed difficult to say a great deal more about the configuration space with any kind of generality. Therefore, we assume that $\ell^2(\mathbf{u})$ is an analytic function of $\mathbf{u}$ and that $E \le V d$. With these assumptions, the Jacobian matrix, whose components are
\begin{equation}
    J_{\alpha i}(\mathbf{u}) = \frac{\partial \ell^2_\alpha(\mathbf{u})}{\partial u_i},
\end{equation}
provides critical information about the mechanism.
Naively, one would expect the configuration space of the mechanism to be $D = M - E$ (for $M > E$). Indeed, the inverse function theorem implies that the configuration space is a smooth $D$ dimensional manifold in any open set of $\mathcal{M}$ in which the Jacobian matrix is full rank.
At such a configuration $\mathbf{u}$, the tangent space coincides with the right null space of $J_{\alpha i}(\mathbf{u})$,
\begin{equation}
\label{eq:zeromodes}
    \sum_i J_{\alpha i}(\mathbf{u}) \delta u_i = 0.
\end{equation}
The solutions $\delta u_i$ of Eq. (\ref{eq:zeromodes}) are called zero modes.

Any point $\mathbf{u}_C$ at which the Jacobian fails to be full rank, on the other hand, we call a critical point, and the corresponding edge lengths $\ell^2_\alpha(\mathbf{u}_C)$ we call a critical value. Critical points are characterized by self stresses, $\sigma_\alpha$, which are elements of the left null space of $J_{\alpha i}(\mathbf{u}_C)$,
\begin{equation}
\label{eq:selfstresses}
    \sum_\alpha \sigma_\alpha J_{\alpha i}(\mathbf{u}_C) = 0.
\end{equation}
Because of their relation to critical points, we will see that self stresses play an important role in the topology of the configuration space.

Sard's theorem ensures that critical values (but not necessarily critical points) are a set of measure zero. In that sense, most choices of edge lengths lead to a configuration space that is a smooth $D$ dimensional manifold. Consequently, any change in the configuration space's topology that occurs as the $L_\alpha$ change must happen at a critical point. Thus, these critical points also govern the overall topology of the configuration space of a mechanism.

In the next section, we proceed to analyze the geometry of the configuration space at and near such critical points.

\subsection{Shape of the configuration space at critical points}
\label{sec:shape}

To understand the shape of the configuration space, we expand $\ell^2_\alpha(\mathbf{u}+\delta \mathbf{u})$ for small deformations, $\delta \mathbf{u}$ having components $\delta u_i$, around the critical point, obtaining
\begin{equation}
\label{eq:criticalpointexpansion}
    0 = \sum_i J_{\alpha i} \delta u_i + \frac{1}{2} \sum_{i j} \frac{\partial^2 \ell^2_\alpha(\mathbf{u}_C)}{\partial u_i \partial u_j} \delta u_i \delta u_j + \mathcal{O}(\delta u^3).
\end{equation}
It is common at this stage to write a formal series expansion, $\delta \mathbf{u} = \delta \mathbf{u}^{(1)} + \delta \mathbf{u}^{(2)} + \cdots$, and substitute it into Eq. (\ref{eq:criticalpointexpansion}). One finds $\delta u^{(1)}$ is a zero mode of the Jacobian satisfying \cite{connelly1996second}
\begin{equation}
\label{eq:secondorderrigidity}
   \frac{1}{2}\sum_\alpha \sum_{i j} \sigma^{(n)}_\alpha \frac{\partial^2 \ell^2_\alpha(\mathbf{u}_C)}{\partial u_i \partial u_j} \delta u_i^{(1)} \delta u_j^{(1)} = 0,
\end{equation}
where $\{ \sigma^{(1)}_\alpha, \sigma^{(2)}_\alpha , \cdots \}$ is a basis for the space of self stresses at $\mathbf{u}_C$.

To proceed, we make further assumptions. The most important of these is that \textit{Eq. (\ref{eq:secondorderrigidity}) completely characterizes the local geometry of the critical point}. It is well-known that if no solution to Eq. (\ref{eq:secondorderrigidity}) exists then the linkage is rigid, but the converse does not necessarily hold. There are mechanisms whose rigidity is only visible at higher order, as well as mechanisms that are rigid at order larger than two but, nevertheless, are mobile \cite{connelly1994higher}. Experience suggests that these examples are rarer than the better behaved examples we consider here, but we are unaware of any results quantifying their rarity or even a simple means to determine when Eq. (\ref{eq:secondorderrigidity}) is sufficient to describe the geometry of the critical point accurately. For the scope of this paper, it will prove sufficient to assume we can safely truncate our expansion of $\delta \mathbf{u}$ at second order and check, \textit{post hoc}, that the results produced by our design procedure satisfy our assumptions.

We will make three other assumptions as well:
\begin{itemize}
    \item[1] \textit{All critical points, $\mathbf{u}_C$, lying on a configuration space of constant $L_\alpha$ are isolated}. There are linkages for which this fails and for which the entire configuration space lies along a sequence of critical points (see, for example, \cite{schicho2022and}). Note, however, that there are also mechanisms with large $D$ which do satisfy this assumption \cite{chen2018branches, kapovich1995moduli, kapovich1999moduli}. In this paper, we will ultimately focus on example mechanisms with only a single degree of freedom, so this will not prove a particularly strong assumption, but in this section we allow $D$ to be general and only specialize to $D=1$ subsequently.

    \item[2] \textit{All critical points have exactly one self stress.} This assumption is certainly not always true. It fails, for example, in flat origami mechanisms \cite{chen2018branches}. Generally, however, we will see that, qualitatively, critical points with several self stresses appear to require more fine-tuning. This assumption implies that there will be $D + 1$ zero modes at each critical point by the rank-nullity theorem applied to the Jacobian matrix at $\mathbf{u}_C$.

    \item[3] \textit{The matrix}
    \begin{eqnarray}\nonumber
        \sum_\alpha \sigma_\alpha \frac{\partial^2 \ell^2_\alpha(\mathbf{u}_C)}{\partial u_i \partial u_j}
    \end{eqnarray}
    \textit{has nonzero eigenvalues when restricted to the zero modes at $\mathbf{u}_C$}. This assumption allows us to simplify the characterization of the critical points. Notice that without assumption 2, this characterization would be more difficult because the Eq. (\ref{eq:secondorderrigidity}) would yield a system of quadratic equations rather than a single equation.
\end{itemize}
While all of these assumptions will play a role in our analysis, one could relax some of them at the expense of complicating the design procedure. Our examples will satisfy them, however, and we leave it for future work to understand which are truly required and which are conveniences.

\begin{figure}
\centering
\includegraphics[width=.8\linewidth]{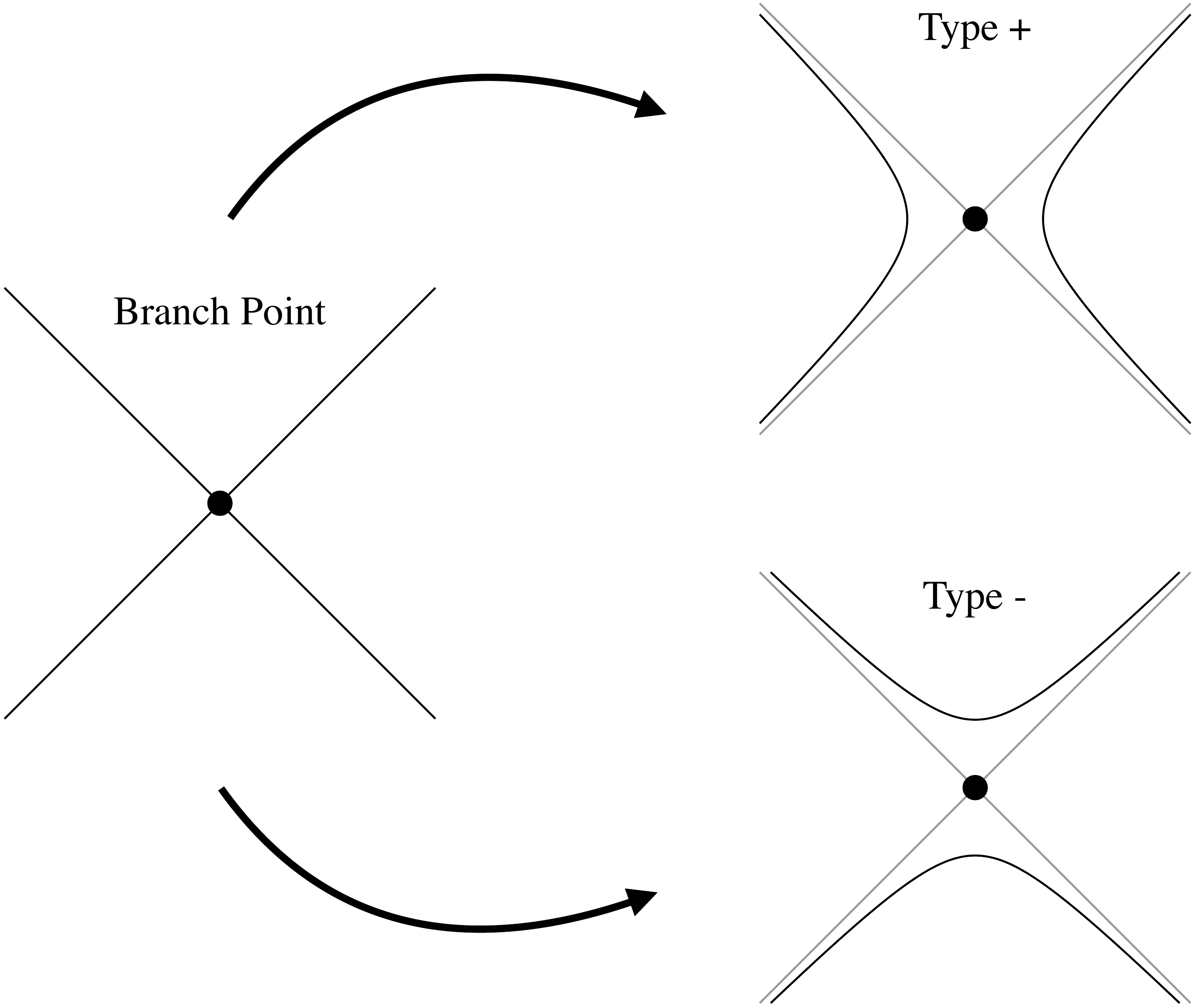}
\caption{Schematic of how a configuration space with a branch point split into one of two types of smooth, disconnected configuration spaces. The choice of sign is arbitrary.}
\label{fig:branch}
\end{figure}

Suppose we choose a basis for the zero modes at $\mathbf{u}_C$, $\{ \boldsymbol{\zeta}_1, \cdots, \boldsymbol{\zeta}_{D+1} \}$, writing $\delta u_i^{(1)} = \sum_n c_n \zeta_{n,i}$. Then Eq. (\ref{eq:secondorderrigidity}) becomes
\begin{equation}
\label{eq:quadraticeq}
	\sum_{nm} \mathcal{Q}_{nm} c_n c_m = 0
\end{equation}
where $\mathcal{Q}_{nm}$ is a symmetric matrix given by
\begin{equation}
	\mathcal{Q}_{n m} = \sum_{i j} \sum_\alpha \zeta_{n,i} \zeta_{m,j}  \sigma^{(1)}_\alpha {\partial^2 \ell^2_\alpha(\mathbf{u}_C)}/{\partial u_i \partial u_j}.
\end{equation}
Under our assumptions, there are just two possibilities. If $\mathcal{Q}_{nm}$ is either positive- or negative-definite, the linkage is rigid: there is no solution to Eq. (\ref{eq:quadraticeq}) other than $c_n = 0$. If $\mathcal{Q}_{nm}$ has a combination of positive and negative eigenvalues, however, the geometry of the configuration space at $\mathbf{u}_C$ is that of a cone. This is precisely what happens in single-vertex flat origami \cite{chen2018branches, berry2020topological} (Fig. \ref{fig:branch}). We call such a point $\mathbf{u}_C$ a branch point, though this space of possible zero modes is sometimes called a kinematic tangent cone \cite{muller2019kinematic}.

\subsection{Shape of the configuration space near critical points}
We next ask what happens to the configuration space of a mechanism when the lengths are deformed from their critical values, $L_\alpha = L_{\alpha}^{(c)} + \delta L_\alpha$. A lengthy calculation shows (see Appendix \ref{sec:quadraticform})
\begin{equation}
\label{eq:conic}
	\sum_{n m} \mathcal{Q}_{n m} (c_n - \delta c_n) (c_m - \delta c_m) = \Delta
\end{equation}
where the deformation is along the zero modes at $\mathbf{u}_C$, $\sum_n c_n \zeta_{n,i}$ as before, and $\delta c_n$ and $\Delta$ are quantities whose value depends linearly on the length changes, $\delta L_\alpha$, to lowest order.

We first consider what happens when $\Delta = 0$. In that case, when $\delta c_n = 0$, Eq. (\ref{eq:conic}) recovers the results from the previous section: there is either a rigid point or a branch point at $c_n = 0$ corresponding to the critical point $\mathbf{u}_C$. When $\delta c_n \ne 0$, however, the critical point itself moves by $\approx \sum_n \delta c_n \zeta_{n,i}$. 

When $\Delta \ne 0$ and $\mathcal{Q}$ has only positive eigenvalues (the critical point is second order rigid), we have two possibilities: (1) $\Delta > 0$ implies the solution to Eq. (\ref{eq:conic}) is an ellipsoid in $D+1$ dimensions (it is almost rigid \cite{holmes2021almost}), and (2) $\Delta < 0$ implies there is no solution to Eq. (\ref{eq:conic}). The opposite occurs if $\mathcal{Q}_{nm}$ has only negative eigenvalues.

Finally, we consider the case of a branch point, for which $\mathcal{Q}_{nm}$ has eigenvalues of opposite sign. 
To develop intuition, it is useful to consider the special case of a branch point when $D=1$. Then $\mathcal{Q}_{nm}$ is a $2 \times 2$ matrix with two eigenvalues of opposite sign. The solutions to Eq. (\ref{eq:conic}) take the form of two hyperbolas in the plane spanned by the zero modes at $\mathbf{u}_C$ whose precise configuration depends on the sign of $\Delta$ (Fig. \ref{fig:branch} for characteristic examples for both signs of $\Delta$). For $D > 1$, branch points also break up into smooth surfaces but do so, presumably, in a more complex way that depends on the signature of $\mathcal{Q}_{nm}$ (see Ref. \cite{berry2020topological} for an example in origami).

\begin{figure}
\centering
\includegraphics[width=.95\linewidth]{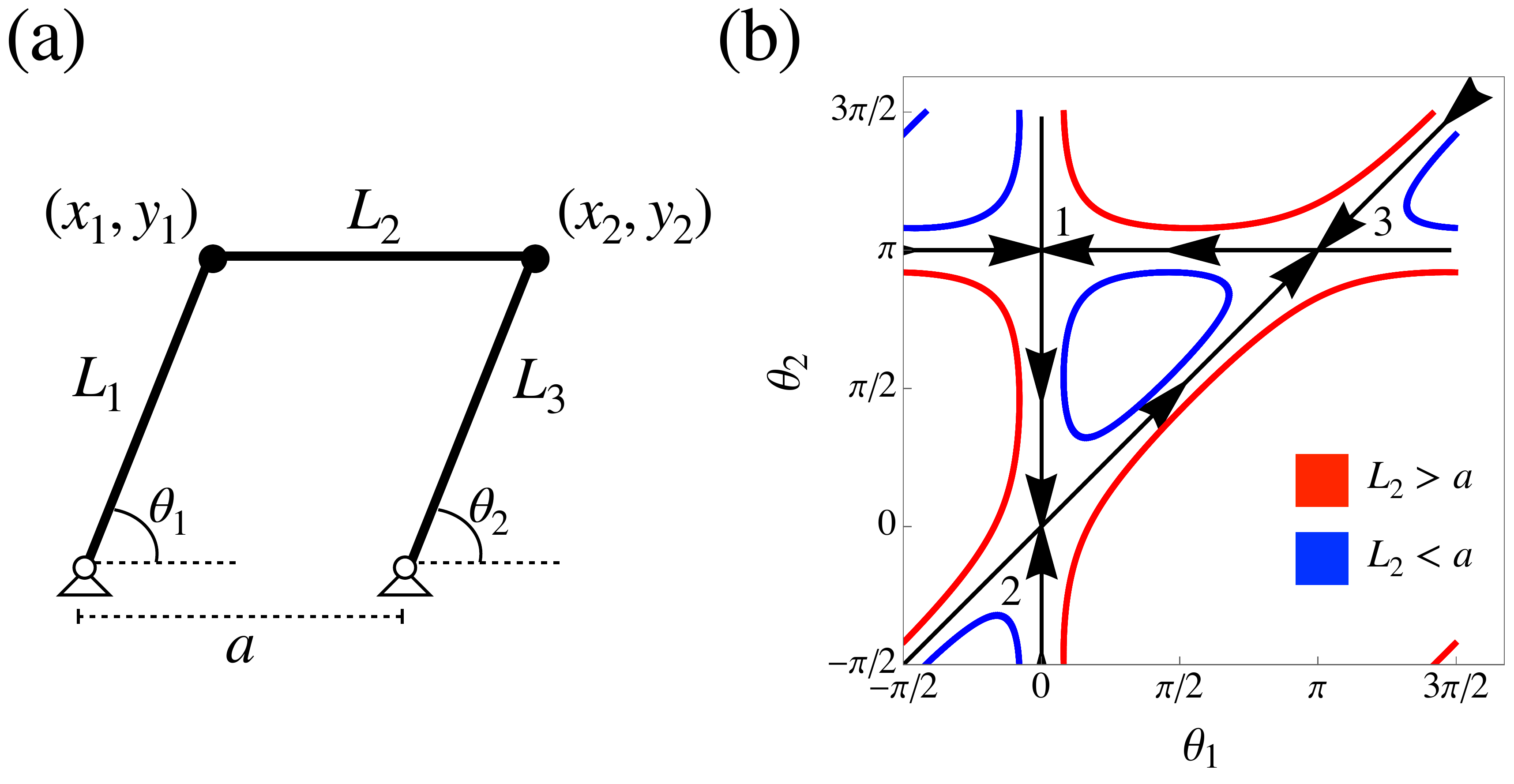}
\caption{(a) Schematic of the planar, four-bar linkage with variables defined. (b) Projection of the configuration space of the two rotor mechanism with $L_1 = L_2 = L_3 = a$ projected into $(\theta_1,\theta_2)$ plane (black). This choice of lengths has three branch-like critical points. Deforming the length of $L_2$ results in a smooth configuration space with either one (red) or two (blue) components. The arrows indicate the direction of the tangent form $t_i(\mathbf{u})$ from Eq. (\ref{eq:tangentfield}).}
\label{fig:one}
\end{figure}

As an illustrative example, we turn to the well-studied four-bar linkage shown in Fig. \ref{fig:one}a. The four-bar linkage is constructed from two rotors of length $L_1$ and $L_3$ pinned at one end and joined at the other by a bar of length $L_2$. The system configuration can be parameterized as a point in four dimensions with coordinates $(x_1,y_1,x_2,y_2)$, and the configuration space is one dimensional. When $L_1 = L_2 = L_3 = a$, there are three branch points each having a single self stress and two zero modes. The configuration space is shown in Fig. \ref{fig:one}b in terms of the two rotor angles $\theta_1$ and $\theta_2$. By slightly increasing the length of $L_2 > a$, the branch points all split into a pair of hyperbolas oriented opposite each other in the quadrants spanned by the configuration space when $L_2 = a$.
On the other hand, $L_2 < a$ results in the branch point splitting into a pair of hyperbolas in the other pair of quadrants. As a result of switching the orientation of the hyperbolas, the configuration space goes from having a single component for $L_2 > a$ to two disconnected components when $L_2 < a$.

\section{Controlling configuration space topology}
\label{sec:controllingtopology}

We noted earlier that the topology of the configuration space cannot change without passing through an intermediate critical point. If it could, this would contradict the notion that the configuration space is smooth when the Jacobian $J_{\alpha i}$ is full rank. This fact and the analysis of Sec. \ref{sec:criticalpoints} suggests a method for controlling the topology of the configuration space: (1) find a set of lengths $L_\alpha$ for which the configuration space has many branch points, and (2) perturb the lengths, $L_\alpha \rightarrow L_\alpha + \delta L_\alpha$, such that the branch points split into smooth hyperbolas in the desired configuration. For the four bar linkage in Fig. \ref{fig:one}b, for example, if we could control how each of the three branch points split independently, we would have complete control over how the configuration space winds around the torus defined by the angles $(\theta_1, \theta_2)$ as well as the number of components in the configuration space.

\subsection{The geometry of the critical configuration set}
Since we are interested in understanding how to choose edge lengths, $L_\alpha$, to control the topology of the configuration space of a linkage, we will consider all possible mechanisms that have the same connectivity but arbitrary values of $L_\alpha$. To do so, we define an antisymmetric tensor
\begin{equation}
\label{eq:tangentfield}
    t_{i_1 \cdots i_D}(\mathbf{u}) = \sum_{j_1 \cdots j_E} \epsilon_{i_1 \cdots i_D j_1 \cdots j_E} \frac{\partial \ell^2_1(\mathbf{u})}{\partial u_{j_1}} \cdots \frac{\partial \ell^2_E(\mathbf{u})}{\partial u_{j_E}}
\end{equation}
where $\epsilon_{i_1 \cdots i_D j_1 \cdots j_E}$ is the antisymmetric Levi-Civita tensor. Importantly, $t_{i_1 \cdots i_D}(\mathbf{u}) = 0$ if and only if $\mathbf{u}$ is a critical point. This is because the components of $t^{i_1 \cdots i_D}$ are the $E \times E$ minors of the Jacobian matrix. When these all vanish the Jacobian matrix has lower rank (see Appendix \ref{sec:tangentform} for a more detailed discussion). Thus, Eq. (\ref{eq:tangentfield}) identifies all possible critical points in mechanisms sharing the same connectivity. Versions of Eq. (\ref{eq:tangentfield}) have been studied to identify singularities in robot manipulators \cite{tchon1995singularity,  tchon1997singularity, lipkin1991, zlatanov1995unifying, donelan2007singularity}.


The tangent form allows us to define the \textit{critical configuration set} as the locus of points for which
\begin{equation}
\label{eq:criticalsetpoints}
    t_{i_1 \cdots i_D}(\mathbf{u}) = 0.
\end{equation}
In many practical cases, and all of the cases we consider in this paper, it is possible to solve Eq. (\ref{eq:criticalsetpoints}) analytically. Note however, that the solutions to Eq. (\ref{eq:criticalsetpoints}) only provide the configurations where the Jacobian of the mechanism is not full rank. Therefore, some of the solutions may not satisfy all of our assumptions from Sec. \ref{sec:shape}. We conjecture that our assumptions are valid on all but a set of measure zero of the critical configuration set but are not aware of or able to produce a proof of this.

To help understand the geometry of the critical configuration set, we return to our previous example, the planar, four-bar linkage from Fig. \ref{fig:one}a. In this example, $D = 1$ but $M = 4$ since the mechanism configurations are specified by points $(x_1, y_1, x_2, y_2)$. If the two pinned vertices are located at $(0,0)$ and $(a,0)$ and we restrict $L_\alpha > 0$ (so no bars have zero length), this critical set is described by the two-dimensional manifold of configurations in which all vertices are co-linear, $y_2 = y_3 = 0$.

For one degree of freedom mechanisms ($D=1$), Eq. (\ref{eq:tangentfield}) provides another way of understanding how the configuration space topology changes with changing lengths near a critical point $\mathbf{u}_C$. In that case, $t_i(\mathbf{u})$ is a vector field everywhere tangent to the zero modes of the mechanism, which follows from the simple fact that it is always orthogonal to the constraints (Appendix \ref{sec:tangentform}). Thus $t_i(\mathbf{u})$ can be thought of as a local vector field whose integral curves trace out curves of constant $L_\alpha$. That is, when $t_i(\mathbf{u}) \ne 0$, curves of constant $L_\alpha$ can be parameterized by the solutions
\begin{equation}
\label{eq:dynamical}
    \frac{d u_i(s)}{ds} = t_i[\mathbf{u}(s)].
\end{equation}
We show this in Fig. \ref{fig:one}b using arrows pointing along $t_i$ projected onto the rotor angles. Because $t_i(\mathbf{u})$ is divergence-free (Appendix \ref{sec:tangentform}), each branch point has two arrows pointing in and two arrows pointing out.
Note that $t_i(\mathbf{u})$ provides a way to think about the mechanism configuration space as a dynamical system. This dynamical system should not be confused with the motions of the physical mechanism, however, which can move either parallel or antiparallel to $t_i(\mathbf{u})$ equally well. This is also distinct from the dynamical system approach obtained for a periodic (or nearly periodic) mechanism as an iterated map \cite{kim2022nonlinear, kim2019conformational}.

Eq. (\ref{eq:dynamical}) also provides an intuitive way to understand the hyperbolas formed by the configuration space near branch points that arise from Eq (\ref{eq:conic}). We project $t_i(\mathbf{u})$ near $\mathbf{u}_C$ onto the plane spanned by the two zero modes, $\boldsymbol{\zeta_1}$ and $\boldsymbol{\zeta_2}$. Since $t_i(\mathbf{u})$ is tangent to the configuration spaces, we expect the trajectories approach this plane as they approach $\mathbf{u}_C$. After projection, we obtain a 2D vector field whose components are,
\begin{equation}
    T_n(c_1, c_2) = \sum_i \zeta_{n,i} t_i(\mathbf{u}_C + c_1 \boldsymbol{\zeta_1}+ c_2 \boldsymbol{\zeta_2}).
\end{equation}
The integral curves of $T_n$ then trace the projection of the configuration space onto the plane spanned by the zero modes near the branch point.

In this projection, the constant $L_\alpha$ trajectories are quite limited in how they can appear. We know that $T_n(0,0) = 0$, but because we assume branch points are isolated, the projected tangent vector $T_n(c_1, c_2) \ne 0$ elsewhere.
Now suppose that the critical point is a branch point. The projection of the configuration space on the plane of zero modes will have the form of a hyperbolic fixed point, with a stable and unstable manifold associated with the configuration space branches that solve Eq. (\ref{eq:quadraticeq}) (Fig. \ref{fig:branch}). Thus, we would generically expect the trajectories near the branch point to appear hyperbolic when projected onto the plane of zero modes.
Though we do not work with second order rigid points here, these considerations also limit what the trajectories do near such rigid points \cite{holmes2021almost}.

\subsection{The geometry of the critical value set}
\begin{figure}
\centering
\includegraphics[width=.8\linewidth]{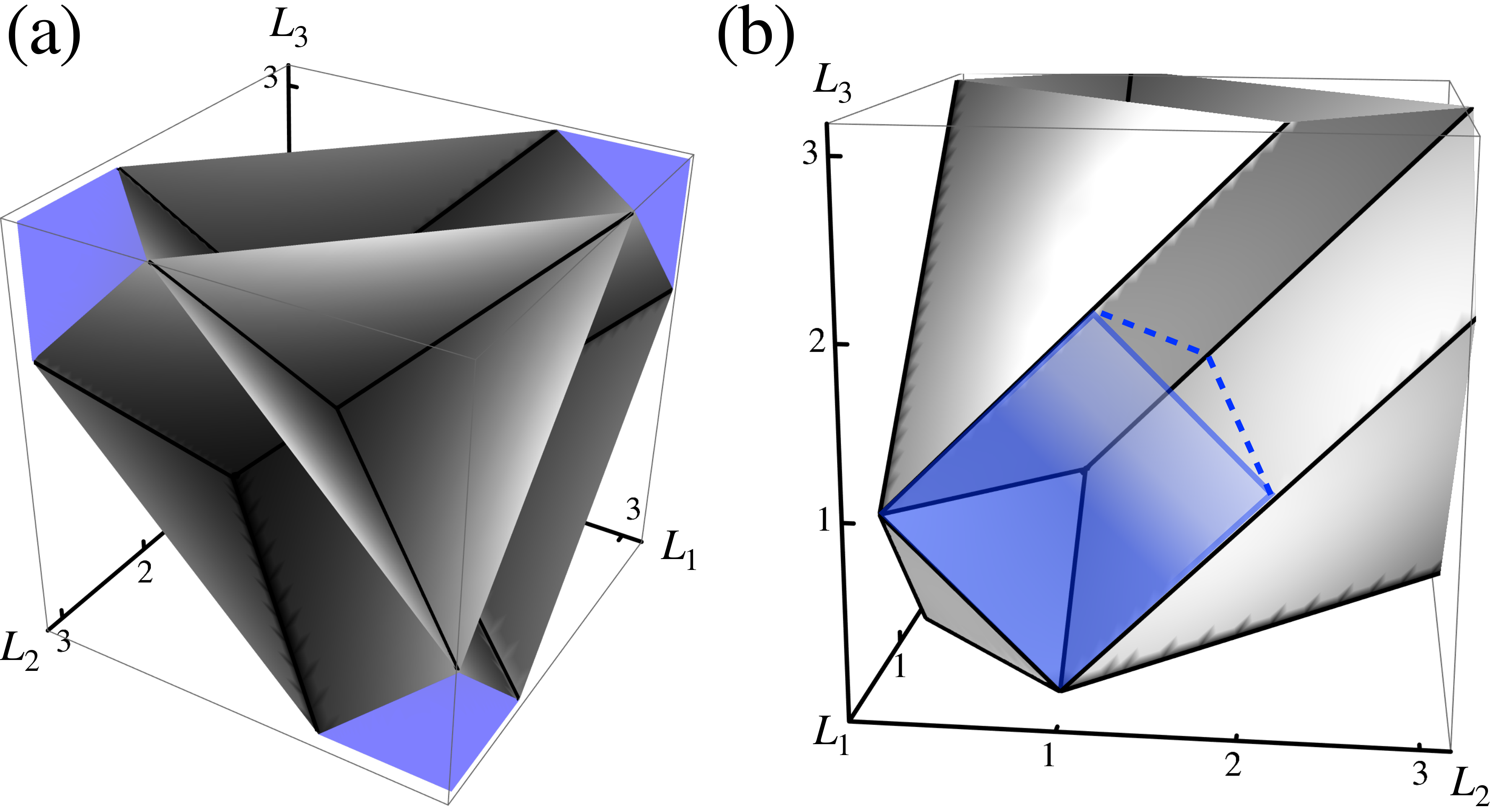}
\caption{The critical value set for the four bar linkage, plotted in terms of $(L_1, L_2, L_3)$ in units of $a$. There is one critical point in the configuration space along any smooth portion of the set. Self intersecting lines indicate choices with two critical points and the triply self intersection point at $(L_1, L_2, L_3) = (a,a,a)$ is the unique choice with three critical points. (b) shows a different view of the surface with a cutout on the $L_1=0$ plane showing the shape of one of the enclosed volumes. }
\label{fig:criticalvalues}
\end{figure}

For any point $\mathbf{u}_C$ in the critical configuration set, $\ell^2_\alpha(\mathbf{u}_C)$ gives its corresponding critical value: the set of squared bar lengths that would be required for the system to be in configuration $\mathbf{u}_C$. We will call the image of the critical configuration set in the space of squared lengths the \textit{critical value set}. We again illustrate with the four-bar linkage: the set of critical values is a self-intersecting surface $(L_1^2, L_2^2, L_3^2) = (x_1^2, (x_2-x_1)^2, (a-x_2)^2  )$. In Fig. \ref{fig:criticalvalues}, we show the critical value set in terms of $(L_1, L_2, L_3)$ rather than the squared lengths to make the surface slightly more compact and easier to understand visually. Note that we have included additional leaves on either the $L_1 = 0$, $L_2 = 0$, or $L_3 = 0$ plane which happen to contain only rigid critical points; this is a natural consequence of the fact that any mechanism with two pinned vertices and one edge having zero length must always be rigid.

Were we to choose the $L_\alpha$ to lie anywhere along the portion of the critical value set in Fig. \ref{fig:criticalvalues}, the resulting linkage would have one or more critical points. It is also apparent that Fig. \ref{fig:criticalvalues} self intersects. At such a self-intersection, there will be multiple critical configurations, $\mathbf{u}_C$, corresponding to the same choices of edge lengths, $L_\alpha$. Thus, if we choose the $L_\alpha$ along a line of self-intersection, there are two branch points. If we choose $L_\alpha$ along a smooth portion of the critical value set, there is only one critical point. Interestingly, Fig. \ref{fig:criticalvalues} shows that at $(L_1, L_2, L_3) = (a,a,a)$ three individual sheets self intersect. Therefore we expect that that choice of $L_\alpha$ is the unique place where three branch points coincide (as seen in Fig. \ref{fig:one}b).

The critical value set contains more information than just the location of critical values. If the critical value set is locally a smooth manifold, the self-stresses at such a critical point are always normal to the critical values (see Sec. \ref{sec:tangentform}). Though the converse is not generally true -- normals need not also be self stresses -- if the critical value set is a manifold of codimension one it must necessarily coincide with the single self stress at that point and there can be no other self stresses. We also see that splitting a branch point amounts to choosing $\delta L_\alpha$ transverse to the critical value set. On one side of the surface in Fig. \ref{fig:criticalvalues}, a branch point splits into one pair of smooth branches; on the other side it splits into the opposite pair. This endows the calculation of how branch points split under small perturbations of the lengths with a concise geometrical meaning.

With this understanding of the critical value set, we can classify the distinct configuration spaces of the four-bar linkage in terms of the $2^3 = 8$ individual volumes enclosed by the surface in Fig. \ref{fig:criticalvalues}. For completeness, we note that these volumes correspond to standard results for the four-bar linkage found in the engineering literature, which can be classified by the sign of three functions \cite{mccarthy2010geometric}
\begin{equation}
\label{eq:configspacelabel}
    \begin{aligned}
        \tau_1 &=a-L_1+L_2-L_3\\
        \tau_2 &=a-L_1-L_2+L_3\\
        \tau_3 &=L_2+L_3-a-L_1
    \end{aligned}
\end{equation}
derived from limits on the angles $\theta_1$ and $\theta_2$. When one of the $\tau_i$ are equal to zero, the configuration space contains the corresponding critical point from Fig. \ref{fig:one}b. Thus,the critical value set in Fig. \ref{fig:criticalvalues} agrees with the surfaces computed in Ref. \cite{barker1985complete, muller1996novel}.

\subsection{Three rotor system}
\label{sec:threerotor}

\begin{figure}
\centering
\includegraphics[width=0.9 \linewidth]{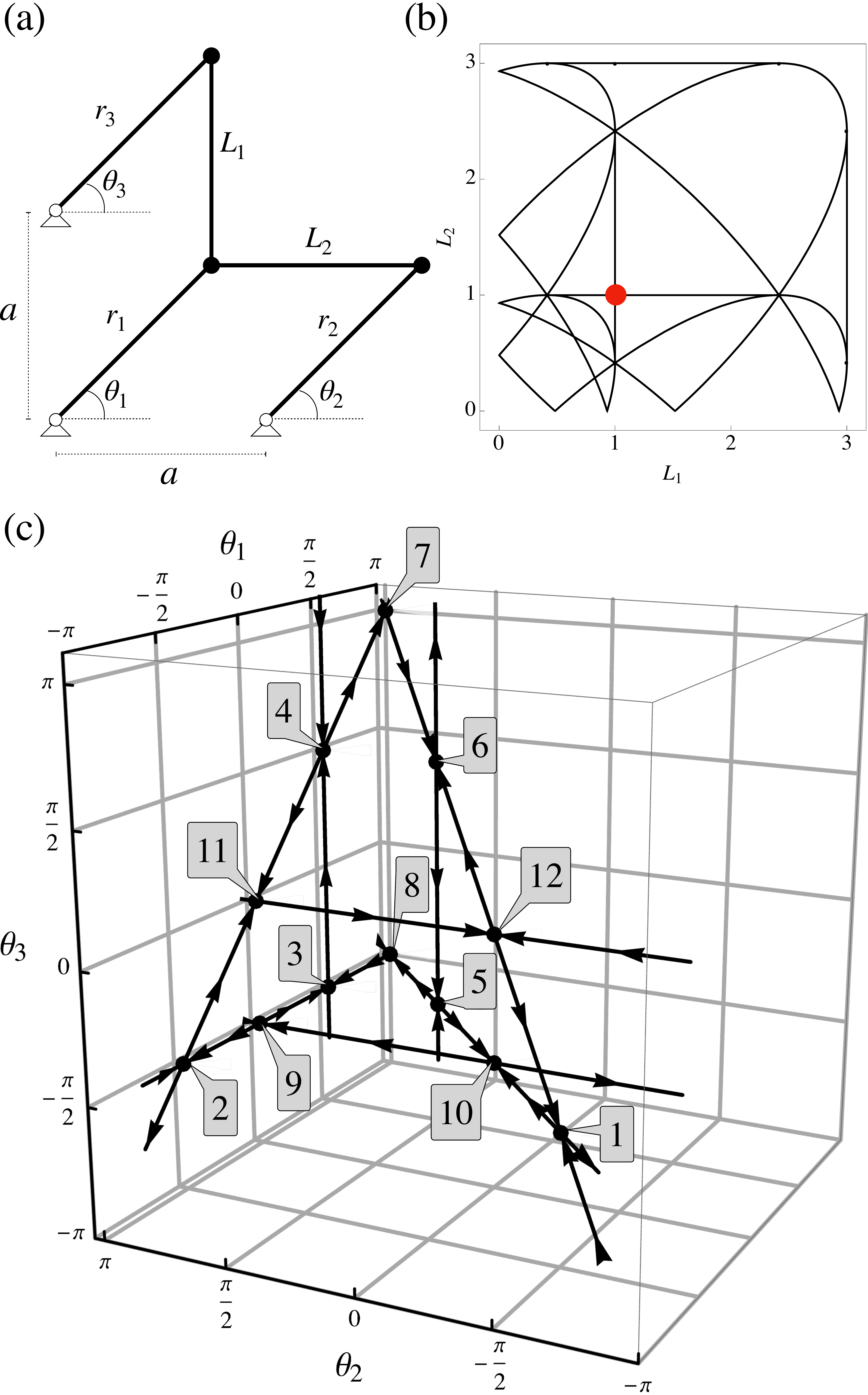}
\caption{(a) Schematic of the planar, three rotor linkage with variables defined. (b) A cross-section of the critical value set for $r_1 = r_2 = r_3 = a$. (c) The 3D configuration space of the three rotor linkage with $r_1 = r_2 = r_3 = L_{13} = L_{2 3} = a$, corresponding to the red point in (b), contains twelve individual critical points. Arrows indicate the orientation of each configuration space segment.}
\label{fig:two}
\end{figure}

Finally, in this section we will use these considerations to describe a design procedure for configuration space topology. To be concrete, it is helpful to consider a specific example, the three-rotor linkage in Fig. \ref{fig:two}a. The three rotor linkage has three pinned joints attached to three bars of length $r_1$, $r_2$, and $r_3$ (the rotors) and whose opposite ends are joined by bars of length $L_1$ and $L_2$. Therefore, $\mathbf{u}$ is a six component vector and the five bars provide constraints, $\ell_\alpha^2(\mathbf{u}) = L_\alpha^2$, that limit the configuration space to a single degree of freedom generically.

Since the mechanism has five bars, it is difficult to visualize the critical set and critical value set. Nevertheless, we can still gain insight by restricting ourselves to the cross-section of $\mathcal{M}$ for which $r_1 = r_2 =r_3 = a$. We plot the cross section of the critical value set with the $(L_1, L_2)$ plane in Fig. \ref{fig:two}b.
While this is a cross-section, the open regions in Fig. \ref{fig:two}b still correspond to structures with different configuration space topologies, with the transitions from one distinct region to another through the critical value set occurring through a branch point. However, it is still a cross section of a higher dimensional space and care must be taken when interpreting the intersections of the critical value set. Choosing $L_1 = L_2 = a$ leads to a configuration space with twelve interconnected branch points, though it appears that only two lines meet at $L_1 = L_2 = a$ in Fig. \ref{fig:two}b. The proliferation of branch points in this example can be understood from the fact that this linkage contains two pairs of four-bar linkages. Choosing all bars to have length $a$, therefore, maximizes the branch points of each individual sub-mechanism.

To identify these branch points, we solve $t_i(\mathbf{u}) = 0$ subject to the length constraints $\ell^2_\alpha(\mathbf{u}) = L_\alpha^2$ using Mathematica (Wolfram). At each critical point, we then solve Eq. (\ref{eq:quadraticeq}) to obtain the tangents to the configuration space. The trajectories in Fig. \ref{fig:two}c are obtained by first stepping along one of the obtained tangent vectors, then stepping along the configuration space in the direction indicated by $t_{i}(\mathbf{u})$ with a step size proportional to its magnitude. The step size is adjusted to maintain the edge lengths to less than one percent strain. Finally, the integration for each segment is terminated when the magnitude of $t_{i}(\mathbf{u})$ falls below a critical threshold, indicating that the integration has reached a point close to the next critical point. Once terminated, we minimize $\sum_i t_i^2$ with respect to the configuration to verify that the integration has found the next branch point. The directions of integration inherited from $t_i(\mathbf{u})$ are indicated by the arrows in Fig. \ref{fig:two}c.

Note, however, that the branch points shown in Fig. \ref{fig:two}c are not all independent. Projecting the configuration space onto the $\theta_1$-$\theta_2$ plane must give the configuration space of the equivalent four-bar linkage found by ignoring the third rotor. In contrast, removing the first rotor is equivalent to the projection onto the $\theta_3$-$\theta_2$ plane. Consequently, any branch points that overlap in one of these two projections must, after a deformation, still be identical in projection and such overlapping branches appear or disappear together. From Fig. \ref{fig:two}, this implies that branch points are paired $\{ (1,2) , (3,5), (4,6), (7,8), (9,11), (10,12) \}$. 

\begin{figure}
\centering
\includegraphics[width=.8\linewidth]{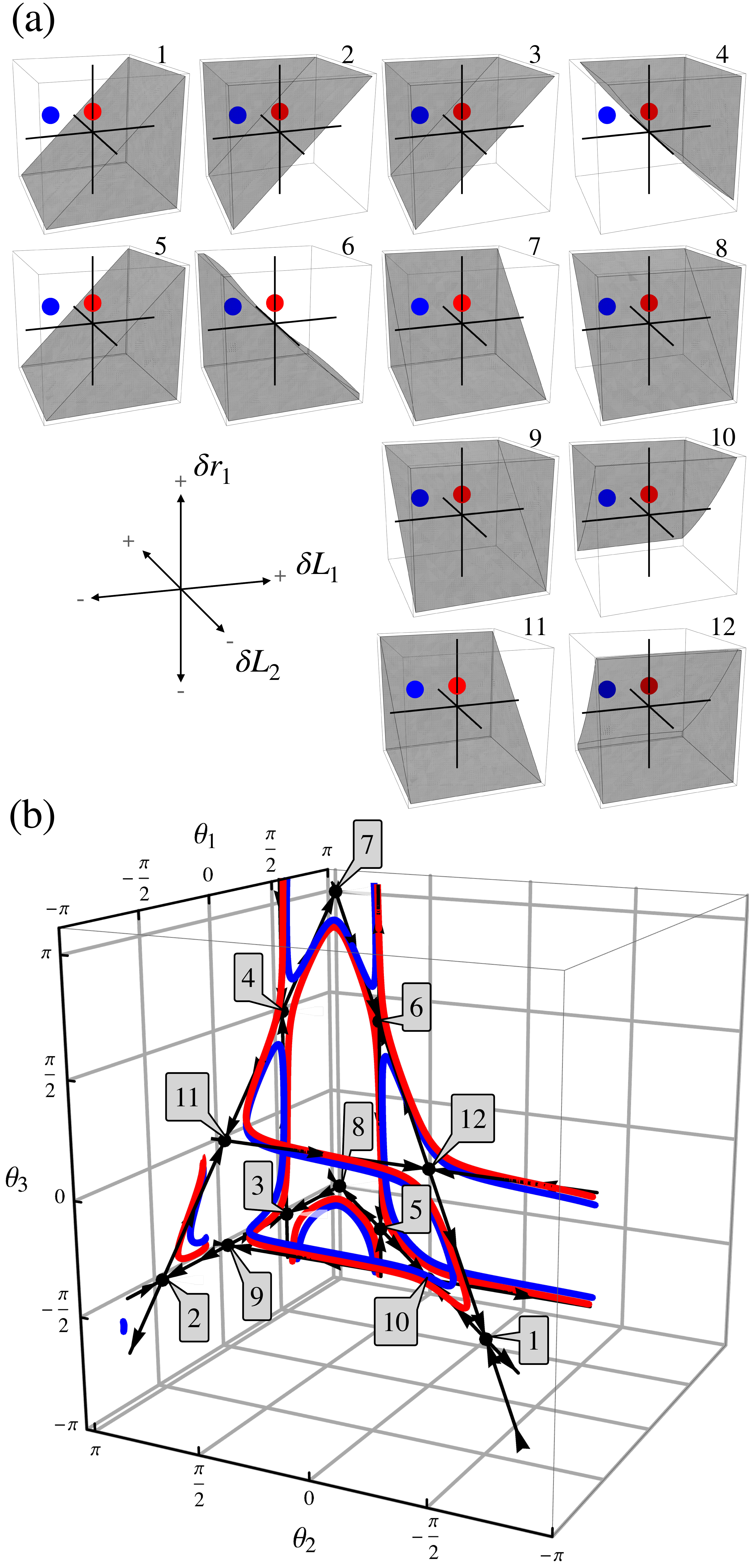}
\caption{A schematic of programming the three rotor system. (a) A map showing how changing the length of $r_1$, $L_1$, and $L_2$ leads to different ways to split the branch points from Fig. \ref{fig:two}. The red dot, corresponding to a change in length $r_1 = 1.05 a$, leads to the red curve in (b). In order to change the topology of the configuration space by changing how branch points 4 and 6 split, an addition change to $L_1 = 0.9 a$ can be effected (blue dot). The new configuration space is shown in (b) as a blue curve.}
\label{fig:programming}
\end{figure}

We finally consider how to ``program'' the configuration space by adjusting the lengths of $r_1$, $L_1$ and $L_2$ away from their critical values. For each critical point, we plot the domain over which $\Delta > 0$ in Fig \ref{fig:programming} as a table for each branch point. We next choose lengths according to the red dot in Fig. \ref{fig:programming}a, which increases $r_1$ at constant $L_1$ and $L_2$. The resulting configuration space is shown as the red curve in Fig. \ref{fig:programming}b. Note that the red point was chosen so that the configuration space is smooth but passes near the branch points. If the red curve has the topology we want already, we can stop now. If we instead wanted to switch the sign of the branch point pair $(4,6)$ to obtain a particular configuration space topology. From Fig. \ref{fig:programming}a we see that the three lengths $(r_1, L_1, L_2)$ distinguish this pair of branch points from the rest. Inspection of Fig. \ref{fig:programming}a suggests an additional change in $L_1$ would switch the way only those two branch points split. The result of this perturbation is the blue curve in Fig. \ref{fig:programming}b. Note that Fig. \ref{fig:programming}b shows that, since each branch point has one self stress, the hyperbolas approach the plane spanned by the two zero modes as expected.

If we limit ourselves to perturbing only the bar lengths $(r_1, L_1, L_2)$, Fig. \ref{fig:programming}a shows even more redundancy in how the branch points split than expected from our previous analysis that branch points split in pairs. That is, just three control lengths are not sufficient to obtain full control over the way the configuration space splits at the branch points. While it would be difficult to plot Fig. \ref{fig:programming} using all five bar lengths, there seems to be no mathematical obstacle to generalizing the analysis to distinguish all six pairs of branch points independently.

\section{The gated Kane-Lubensky chain}
\label{sec:gatedchain}

\begin{figure}
\centering
\includegraphics[width=\linewidth]{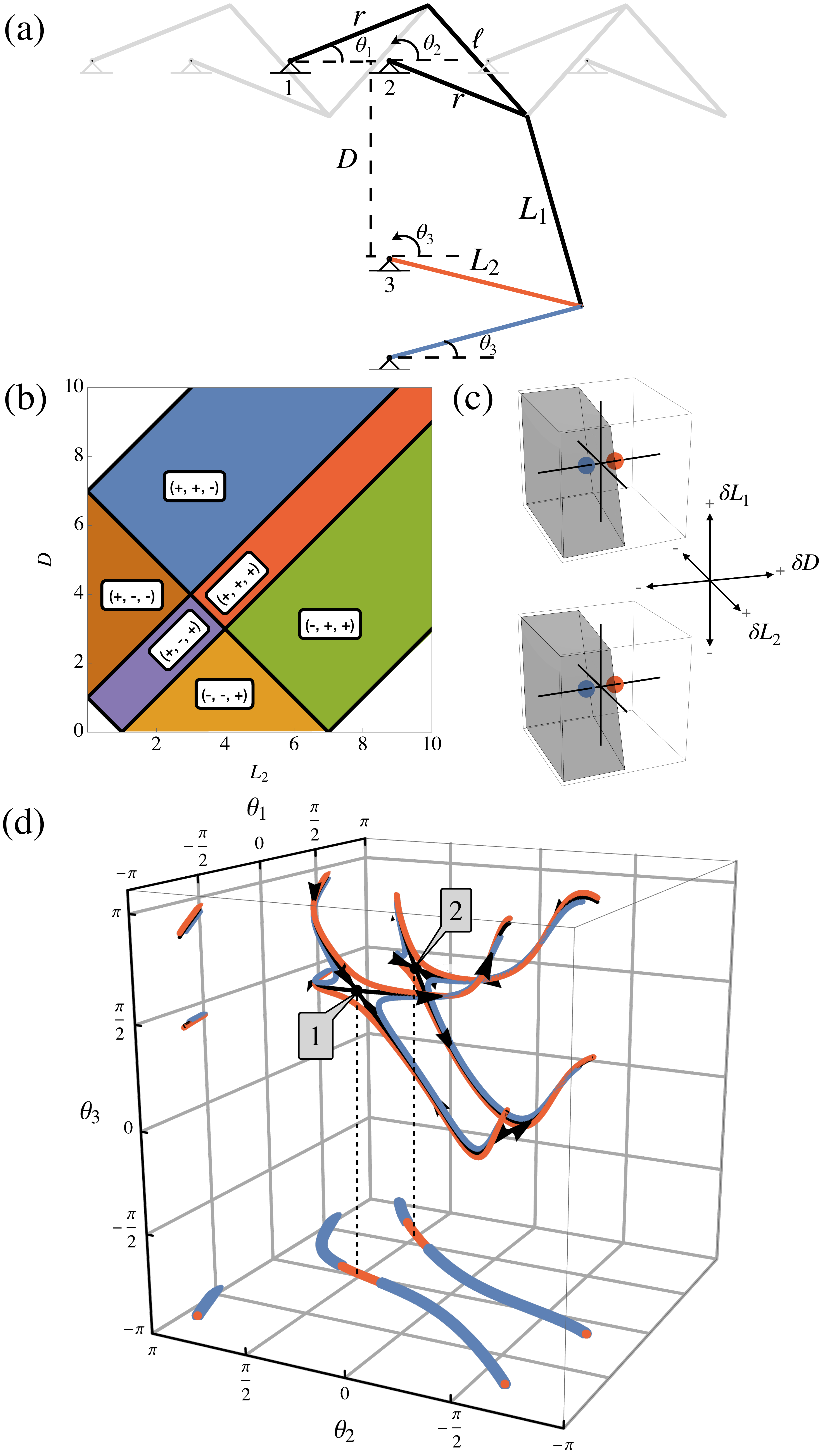}
\caption{(a) A gated and ungated Kane-Lubensky chain controlled by the length $D$. (b) A cross-section of the critical set with $r = \ell = L_1 = 3 a$ and $L_2 = 4 a$. There is a critical point at $L_2 =4 a$ and $L_4 = 5 a$. (c) Changing the position of the third rotor or the lengths of two beams can control whether the chain is gated or ungated. (d) The configuration space at and near the critical point as a function of the three rotor angles, and the projection of that configuration space onto the $\theta_1$-$\theta_2$ plane.}
\label{fig:gate}
\end{figure}

We finally apply our design methodology to design a mechanism that gates the propagation of a soliton in the Kane-Lubensky (KL) chain \cite{kane2014topological}. The KL chain is a topologically polarized lattice of rotors that has a zero mode on either the left edge or the right edge, depending on the choice of bar lengths. It was later discovered that the KL chain actually supported two distinct families of propagating solitons, the ``flipper'' and the ``spinner'' \cite{chen2014nonlinear}, that allowed a continuous pathway between the left and right edge modes. The spinner soliton, however, is topologically protected by the shape of the configuration space \cite{chen2014nonlinear,lo2021topology}. In this section, we modify a single unit cell of a spinner-supporting KL chain with rotor length $r = 3 a/2$ and $\ell = 3 a/2$ by adding an additional two bars and one pinned vertex (Fig. \ref{fig:gate}a). 

In the spinner phase of the KL chain, a full cycle consists of the soliton traveling back and forth across the chain once, and the KL chain returning to its initial configuration. After one full cycle, each rotor in the KL chain has rotated by $2\pi$, with each rotor rotating by $\pi$ each time the soliton passes. Here, we will show that these additional components can act as a gate by opening a gap in the full $2\pi$ rotation of the KL chain rotors, thereby obstructing the passage of the soliton.

In addition to the length of the two additional bars, $L_1$ and $L_2$, we also allow the location of the pinned vertex to be set an arbitrary distance $D$ from the KL chain. In order to allow for different positions of the third pinned vertex, we augment $\mathbf{u}$ to include the $y$ coordinate of the 3rd vertex but also augment the constraint functions to pin that vertex's y position. Thus, we use a constraint map
\begin{equation}
	f_\alpha(\mathbf{u}) = \left( \begin{array}{c}
							\ell_1^2(\mathbf{u})\\
							\vdots\\
							\ell_E^2(\mathbf{u})\\
							D^2(\mathbf{u})
						\end{array} \right),
\end{equation}
where $D(\mathbf{u})$ is the function that determines the distance between pinned vertex $3$ and $2$. Thus, the generalized constraints $f_\alpha(\mathbf{u})$ is a smooth function whose solution allows us to pin vertex $3$ by setting the length $D$ in Fig. \ref{fig:gate}a to an arbitrary value.

\begin{figure*}
\centering
\includegraphics[width=\textwidth]{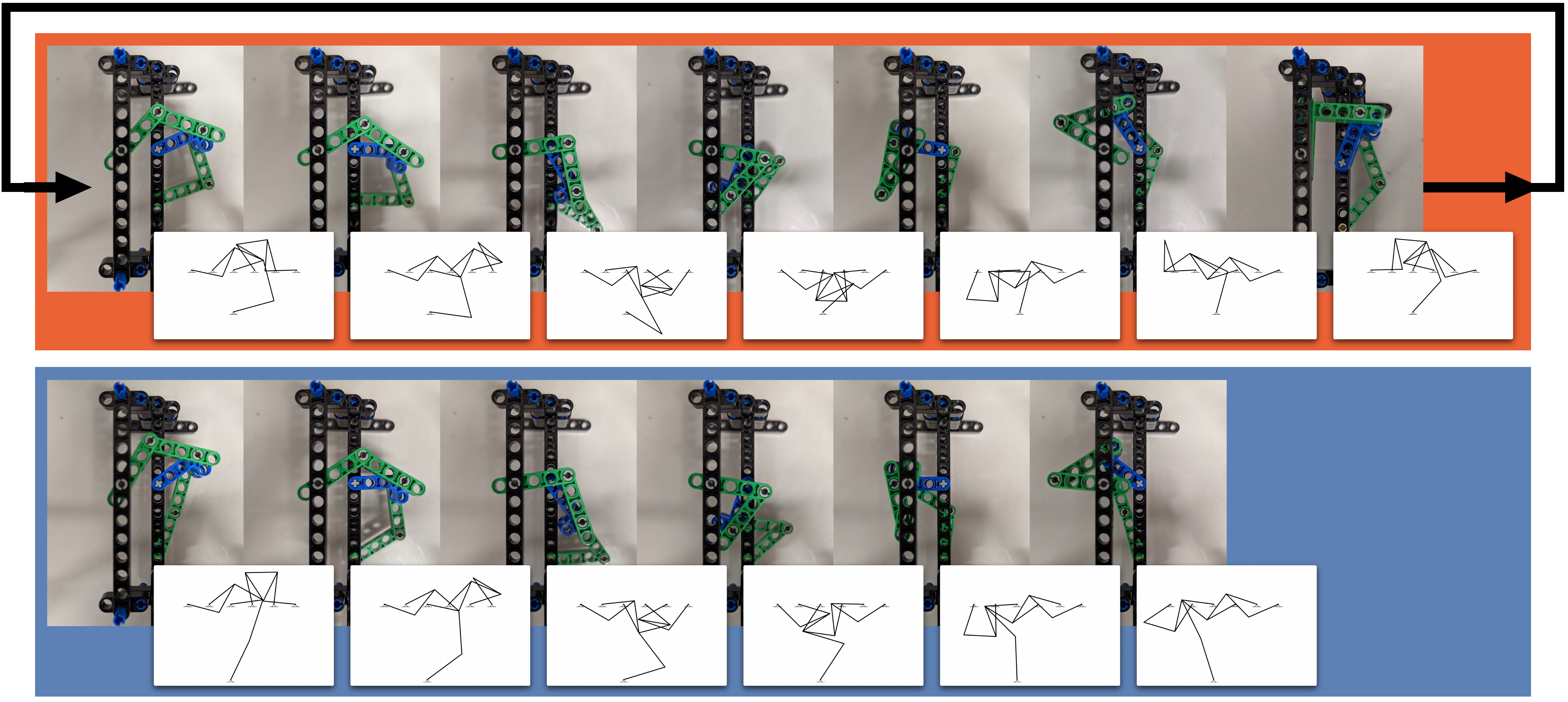}
\caption{Top row (red): Ungated device made from LEGOs with the corresponding simulation. This device can continue rotating and return back to its initial position, as indicated by the arrow. Bottom row (blue): Gated device made from LEGOs with the corresponding simulation. This device gets stuck in the configuration shown in the last frame and is forced to reverse direction in order to continue moving.}
\label{fig:experiments}
\end{figure*}

Using this generalized formulation, we can compute a cross section of the critical value set with $r = \ell = 3 a/2$ and $L_1 = 2 a$ (Fig. \ref{fig:gate}b). 
Fig. \ref{fig:gate}b shows that there are six distinct regions separated by critical points. The labels on each region correspond to the sign of $\tau_1$, $\tau_2$, and $\tau_3$ from Eq. \ref{eq:configspacelabel} with respect to the four-bar linkage between vertices two and three. For concreteness, we choose $L_2 = 2 a$ and $D = 5a/2$, on the boundary between the blue and red regions, as the initial lengths for our gate, resulting in a configuration space with two critical points (Fig. \ref{fig:gate}d). When $D < 5 a/2$, the system is in the ``red'' regime and when $D > 5 a/2$ it is in the ``blue'' regime. This choice determines whether the KL chain rotors wind around fully or not. Note that the projection of the configuration space in the $\theta_1$-$\theta_2$ plane never changes shape, but that the change in how the branch points split into hyperbolas determines whether the full range of angles is accessible to the system or not. To verify that the red and blue regimes correspond to ungated and gated behavior of the KL chain device, we use the \textit{mechanisms} package \footnote{Availabe on https://github.com/cdsantan/mechanisms} in Mathematica (Wolfram) to calculate the infinitesimal motions of the linkage and animate the those motions. As shown in Fig. \ref{fig:experiments}, changing $D$ controls whether or not the soliton can complete a full cycle along the KL chain.

To test our design, we constructed the gated KL chain in Fig. \ref{fig:gate}a numerically and constructed a single unit cell and gate from LEGO\textsuperscript{\texttrademark} pieces. The design of the LEGO gate was chosen to be compatible with the LEGO realization of a KL chain shown in \cite{chen2014nonlinear}. When testing different examples, we pushed on the various bars and rotors in the device to move it through all possible configurations. We tracked how the rotors 1 and 2 moved to determine if the gate was preventing a soliton from propagating. Fig. \ref{fig:experiments} shows a comparison between the simulated and LEGO chains with both $D$ larger and smaller than $5a/2$. Movies of both chains in the gated and ungated states are provided in the Supplementary Material. In the case of an ungated chain, the soliton propagates from one end of the chain to the other (and back); for a gated chain the soliton propagates up to the location of the gate but is reflected. 

Interestingly, the size of the gap in Fig. \ref{fig:gate}d is important for determining how the soliton is reflected from the gate. For very small gaps, which occurs when $D$ is close to its critical value, the soliton can, temporarily, pass the gate but is, ultimately, prevented from completing an entire cycle. For larger gaps, when $D$ is farther from its critical value, the soliton appears to reflect from the gate. From Fig. \ref{fig:gate}c, the same effect can be achieved by changing the size of $L_2$ instead of $D$, since the plane divided the gray region from the transparent region is slightly angled in that direction. Movies of both simulated and LEGO chains that switch between the gated and ungated states by changing $L_2$ are also provided in the Supplementary Material.

Our analysis shows that the presence of a gap in the $(\theta_1, \theta_2)$ plane blocks soliton propagation. In the example of Fig. \ref{fig:gate}a, changing the length $D$ moves the device from from the gated (blue) region to the ungated (red) region of Fig. \ref{fig:gate}b. However, this is not the only pair of regions that produces a functioning gate. Indeed, the regions indicated in Fig. \ref{fig:gate}b as $(+, +, +)$ and $(-, -, +)$ are ungated with respect to propagation of the soliton, whereas the remaining regions are gated. Numerical experiments further show that if we had chosen $L_1$ to change length as well, we would have found even more regions of both gated and ungated behavior as we extended Fig. \ref{fig:gate}b. It becomes clear that there is a great deal of flexibility when choosing $D$, $L_1$, and $L_2$ to produce the desired dynamics of the final KL chain and gate system.

\section{Conclusions}
In this paper, we have described a procedure to design the topology of the configuration space of mechanical linkage. The idea rests on the ability to identify critical points and, especially, branch points -- singular configurations of a linkage in which several pathways meet. By analyzing the shape of the configuration space near these branch points, we are able to design perturbations to the lengths and positions of a fixed set of vertices that change the shape of the topology of the configuration space in well-defined ways. As a demonstration, we used our techniques to design a gate for the propagation of the spinner soliton in a Kane-Lubensky chain. While we applied our approach to linkages with fixed edge length, there is no reason they would not also apply more generally to other systems with holonomic constraints. 

Because the design procedure works by controlling configuration space topology, the resulting mechanisms should be quite robust to fabrication errors and the tolerance of the joints, so long as one chooses lengths $L_\alpha$ sufficiently far from the critical value set.

It would be interesting to extend this work in a few further directions. First, when bars are no longer rigid but elastic, there arises the possibility of a snap through transition between the different hyperbolas on either side of a branch point. Indeed, tuning various branches close to or farther from a branch point could be used to tune the ease of initiating a snap through transition. This could potentially lead to mechanical structures and mechanical metamaterials whose mechanical response can be reprogrammed \textit{in situ}.

A second interesting extension would be to consider mechanisms built from responsive materials that are sensitive to external stimuli. In that case, the dynamic increase or decrease in the lengths of bars could be used to drive the pathway of a mechanism in an environmentally dependent manner. This could also be affected if the positions of certain pinned vertices could be made to depend on the external environment or the state of a second input mechanism. This would enable the realization of simple mechanical logic that is robust to some damage because it relies only on the topology of a configuration space \cite{song2019additively, meng2021bistability}.

Finally, we note that our design principle exploits the fact that the configuration space topology can only change at critical points -- configurations where the Jacobian of the constraints fails to be full rank. Our approach is somewhat reminiscent of Morse theory, in which the extrema of a scalar function can be related to the topology of the space on which that function is defined \cite{nicolaescu2007invitation}. Morse theory has been used to study the configuration spaces of spherical (and other) linkages \cite{kapovich1995moduli, kapovich1999moduli}, but we leave it to future work to make this connection more precise.


\begin{acknowledgements}
We thank useful conversations and several helpful comments on a previous version of this manuscript with Manu Mannatil. We acknowledge funding from the National Science Foundation through grant NSF DMR-1822638 and Graduate Research Fellowship under Grant No. 451512, and funding from the US Army Research Office through grant W911NF-21-1-0068.
\end{acknowledgements}

\appendix

\vspace{1cm}

\section{Perturbing quadratic critical points}
\label{sec:quadraticform}

In this appendix, we will show that Eq. (\ref{eq:conic}) does, indeed, describe the configuration space near a critical point when the lengths of a linkage are perturbed from their critical values. We assume we have a mechanism with $E$ edges and $V$ vertices in $d$ dimensions with $d V > E$. We further suppose that the configuration of the mechanism is at a critical point, $\mathbf{u}_C$, with corresponding critical values $(L_\alpha^{(c)})^2$. Let $\mathbf{u} = \mathbf{u}_C + \delta \mathbf{u}$ and correspondingly $L_\alpha = L_\alpha^{(c)} + \delta L_\alpha$, and expand the squared lengths to quadratic order, using Eq. (\ref{eq:constraints},
\begin{eqnarray}
\label{eq:one}
	2 L_\alpha^{(c)} \delta L_\alpha + \delta L_\alpha^2	&=& \sum_i \frac{\partial \ell_\alpha^2(\mathbf{u}_C) }{\partial u_i} \delta u_i \\
\nonumber									& & + \sum_{i j} \frac{1}{2} \frac{\partial^2 \ell_\alpha^2(\mathbf{u}_C)}{\partial u_i \partial u_j} \delta u_i \delta u_j  
\end{eqnarray}
Finally, as in the main text, we assume that Eq. (\ref{eq:one}) completely characterizes the critical point, and that there is one self stress at $\mathbf{u}_C$, with components $\sigma_\alpha$, and two zero modes, with components $\zeta_{1,i}$ and $\zeta_{2,i}$.

It will prove convenient to express Eq. (\ref{eq:one}) using an orthonormal basis in the space of square lengths, $\{ \sigma_\alpha , e_{\alpha}^{(1)}, \cdots e_\alpha^{(E-1)} \}$. We similarly write $\delta u_i$ in an orthonormal basis $\{ \zeta_{1,i}, \zeta_{2,i}, \eta_{1,i}, \cdots, \eta_{E-1,i} \}$,
\begin{equation}
    \delta u_i = c_1 \zeta_{1,i} + c_2 \zeta_{2,i} + \sum_{I=1}^{E-1} a_I \eta_{I,i}.
\end{equation}

We first contract Eq. (\ref{eq:one}) with $\sigma_\alpha$, we obtain an equation that can be expressed as
\begin{equation}
\label{eq:matrixeq}
   \left( \begin{array}{cc}
        \mathbf{c}^T & \mathbf{a}^T
    \end{array}\right)  \left( \begin{array}{cc}
        \mathcal{Q} & \mathcal{B} \\
        \mathcal{B}^T & \mathcal{M}
    \end{array}\right) \left( \begin{array}{c}
        \mathbf{c}\\
        \mathbf{a}
    \end{array}\right) = \tilde{\Delta},
\end{equation}
where the components of the matrices are given by
\begin{equation}
    \tilde{\Delta} = \sum_\alpha \sigma_\alpha \left( 2 L_\alpha^{(c)} \delta L_\alpha + \delta L_\alpha^2 \right),
\end{equation}
\begin{equation}
    \mathcal{Q}_{nm} = \frac{1}{2} \sum_{\alpha i j} \zeta_{n,i}\zeta_{m,j} \sigma_\alpha \frac{\partial^2 \ell_\alpha^2}{\partial u_i \partial u_j},
\end{equation}
\begin{equation}
    \mathcal{M}_{nm} = \frac{1}{2} \sum_{\alpha i j} \eta_{n,i}\eta_{m,j} \sigma_\alpha \frac{\partial^2 \ell_\alpha^2}{\partial u_i \partial u_j},
\end{equation}
and
\begin{equation}
    \mathcal{B}_{nm} = \frac{1}{2} \sum_{\alpha i j} \zeta_{n,i}\eta_{m,j} \sigma_\alpha \frac{\partial^2 \ell_\alpha^2}{\partial u_i \partial u_j}.
\end{equation}
We also assume that $a_I$ are the components of the vector $\mathbf{a}$ and that $c_1$ and $c_2$ are the components of a two-dimensional vector $\mathbf{c}$.
Finally, we complete the square in Eq. (\ref{eq:matrixeq}) to obtain
\begin{equation}
\label{eq:conicappendix}
    \left( \mathbf{c} + \mathcal{Q}^{-1} \mathcal{B} \mathbf{a} \right)^T \mathcal{Q} \left( \mathbf{c} + \mathcal{Q}^{-1} \mathcal{B} \mathbf{a} \right) = \tilde{\Delta} - \mathbf{a}^T \mathcal{B}^T \mathcal{Q}^{-1} \mathcal{B} \mathbf{a}.
\end{equation}
Note that $\mathcal{Q}^{-1}$ exists because all of the eigenvalues of $\mathcal{Q}$ are nonzero by assumption.

Already, Eq. (\ref{eq:matrixeq}) is in the form of a conic section whose form depends on the eigenvalues of $\mathcal{Q}$. What remains is to show that $\mathbf{a}$ depends only on the length changes (and not $\mathbf{c}$) to lowest order and, ultimately, to find an expression to determine it.

To do this, we project Eq. (\ref{eq:one}) onto the remaining basis vectors, $e_\alpha^{(n)}$, in the space of square lengths. We obtain
\begin{widetext}
\begin{eqnarray}
\label{eq:linear}
	\sum_m \sum_i \sum_\alpha e_\alpha^{(n)} \frac{\partial \ell_\alpha^2(\mathbf{u}_C) }{\partial u_i} \eta_{m,i} a_m +  \frac{1}{2} \sum_{i j \alpha}  e_\alpha^{(n)} \frac{\partial^2 \ell_\alpha^2(\mathbf{u}_C)}{\partial u_i \partial u_j} \delta u_i \delta u_j &=& \sum_\alpha e_\alpha^{(n)}  \left(2 L_\alpha^{(c)} \delta L_\alpha + \delta L_\alpha^2\right)
\end{eqnarray}
\end{widetext}
There are $E-1$ equations in Eq. (\ref{eq:linear}) and $d V - E+1$ zero modes at the critical point, the space spanned by $\delta u_i^{\perp}$ is $d V - (d V - E + 1) = E - 1$ dimensional. The matrix appearing in Eq. (\ref{eq:linear}) is, consequently, square. Since we have already removed zero modes and self stresses, it is also invertible. We define a new matrix $\mathbf{M}$ such that its inverse $\mathbf{M}^{-1}$ is given by the components,
\begin{equation}
    \mathbf{M}^{-1}_{nm} = \sum_i \sum_\alpha e_\alpha^{(n)} \frac{\partial \ell_\alpha^2(\mathbf{u}_C) }{\partial u_i} \eta_{m,i}.
\end{equation}
This then allows us to solve Eq. (\ref{eq:linear}) in powers of both $\delta L_\alpha$ and $\mathbf{c}$. To first order in both, we obtain
\begin{equation}
    a_n \approx \sum_m M_{n m} \sum_\alpha 2 e^{(m)}_{\alpha} L_\alpha^{(c)} \delta L_\alpha + \mathcal{O}(c \delta L, c^2, \delta L^2).
\end{equation}

We can now put together the results by defining
\begin{equation}
    \delta \mathbf{c} = -\mathcal{Q}^{-1} \mathcal{B} \mathbf{a}
\end{equation}
and
\begin{equation}
    \Delta = \tilde{\Delta} - \delta \mathbf{c} \mathcal{Q} \delta \mathbf{c}
\end{equation}
to obtain
\begin{equation}
\label{eq:conicresult}
    \left( \delta \mathbf{c} - \delta \mathbf{c}  \right)^T \mathcal{Q} \left( \delta \mathbf{c} - \delta \mathbf{c}  \right) = \Delta
\end{equation}
where $\Delta$ and $\delta \mathbf{c}$ depend linearly on the changes in lengths to lowest order. Therefore, small perturbations of the length are seen to produce trajectories that lie on a 2D conic section with a perturbed center.

While this is a rather intricate derivation, we could have obtained the correct answer up to order $\delta \mathbf{u} \sim \delta L^{1/2}$ more simply by assuming $\mathcal{O}(\mathbf{a}) \sim \mathcal{O}(\mathbf{c})$. We have found the full form of Eq. (\ref{eq:conicresult}) to be more useful in perturbing larger linkages, however, as it better captures the case that changes in the bar lengths perturb but do not completely eliminate critical points in the configuration space of a linkage.

\section{Properties of the tangent form}
\label{sec:tangentform}

The tangent form is defined as
\begin{equation}
\label{eq:tangentfield2}
    t^{i_1 \cdots i_D}(\mathbf{u}) = \sum_{j_1 \cdots j_N} \epsilon^{i_1 \cdots i_D j_1 \cdots j_N} \frac{\partial f_1(\mathbf{u})}{\partial u_{j_1}} \cdots \frac{\partial f_N(\mathbf{u})}{\partial u_{j_N}},
\end{equation}
where $\epsilon^{i_1 \cdots i_D j_1 \cdots j_N}$ is the antisymmetric Levi-Civita tensor. Next we compute some simple properties of the tangent form.\\

\noindent \textbf{The tangent form is divergence free.} This can be seen from the following calculation,
\begin{eqnarray}
\label{eq:tangentfield3}
   \frac{\partial t^{i_1 \cdots i_D}(\mathbf{u}) }{\partial u_{i_1}}	&=& \sum_{j_1 \cdots j_N} \epsilon^{i_1 \cdots i_D j_1 \cdots j_N} \frac{\partial^2 f_1(\mathbf{u})}{\partial u_{i_1} \partial u_{j_1}} \cdots \frac{\partial f_N(\mathbf{u})}{\partial u_{j_N}}  \\
\nonumber
	&\cdots &+\sum_{j_1 \cdots j_N} \epsilon^{i_1 \cdots i_D j_1 \cdots j_N}  \frac{\partial f_1(\mathbf{u})}{\partial u_{j_1}} \cdots \frac{\partial}{\partial u_{i_1}} \frac{\partial f_N(\mathbf{u})}{\partial u_{j_N}}\\
\nonumber
	&=& 0
\end{eqnarray}
where each term is zero due to the antisymmetry of the Levi-Civita tensor and the symmetry of partial derivatives.\\

\noindent\textbf{For one degree of freedom mechanisms, the tangent form is a vector tangent to the configuration space away from critical points.} First, we note that
\begin{equation}
	\label{eq:orthog}
	\sum_{i_1} \frac{\partial f_\alpha}{\partial u_{i_1}} t^{i_1 \cdots i_D}(\mathbf{u}(s)) = 0
\end{equation}
which implies that $\frac{\partial f_\alpha}{\partial u_{i}} t^{i}(\mathbf{u}(s)) = 0.$ Now suppose that $\mathbf{u}(s)$ traces the configuration space in a region where $t^{i_1 \cdots i_D}(\mathbf{u}(s))$ is nonzero. Then
\begin{equation}
	\sum_i \frac{ \partial f_\alpha(\mathbf{u}(s))}{\partial u_i} \frac{\partial u_i(s)}{\partial s} = 0.
\end{equation}
Hence the configuration space is perpendicular to all of the $\partial f_\alpha(\mathbf{u})/\partial u_i$ but $t_i(\mathbf{u})$ is also perpendicular to all of them. Hence, they must be parallel. The more general case for mechanisms with more than one degree of freedom is more subtle but can also be computed.\\

\noindent\textbf{The tangent form is zero at $\mathbf{u}$ if and only if $\mathbf{u}$ is a critical point.} Ultimately, this is a consequence of the fact that the components of $t^{i_1 \cdots i_D}(\mathbf{u})$ are the $E \times E$ minors of the Jacobian of $\ell^2(\mathbf{u})$. Nevertheless, we demonstrate it here for completeness. There are $E$ functions
\begin{equation}
	\left\{ \frac{\partial f_1(\mathbf{u})}{\partial u_i}, \cdots , \frac{\partial f_E(\mathbf{u})}{\partial u_i} \right\}.
\end{equation}
Since the zero modes are defined by the nonzero solutions, $\delta u_i$ of
\begin{equation}
	\label{eq:zeromode}
	\sum_i \frac{\partial f_\alpha(\mathbf{u})}{\partial u_i} \delta u_i = 0
\end{equation}
the zero modes are in the orthogonal complement of the span of the vectors $\partial f_\alpha(\mathbf{u})/\partial u_i$. At a critical point, there must be additional zero modes and so the $\partial f_\alpha(\mathbf{u})/\partial u_i$ span a lower dimensional space and can no longer be linearly independent. Without loss of generality, we can take it to be $\alpha =1$ so
\begin{equation}
	\label{eq:lineardependence}
	\frac{\partial f_1(\mathbf{u})}{\partial u_i} = \sum_{\beta > 1} c_\alpha \frac{\partial f_\beta(\mathbf{u})}{\partial u_i}.
\end{equation}
Substituting this into the definition of $t_{i_1 \cdots i_D}(\mathbf{u})$ and using Eq. (\ref{eq:orthog}), we immediately obtain $t_{i_1 \cdots i_D}(\mathbf{u}) = 0$.

Similarly, if $t_{i_1 \cdots i_D}(\mathbf{u}) = 0$ then the $\partial f_\alpha(\mathbf{u})/\partial u_i$ cannot all be linearly independent. One way to do see this is to choose $D$ vectors $\mathbf{v}_n$ orthogonal to the $\partial f_\alpha(\mathbf{u})/\partial u_i$ for all $\alpha$ as well as to each other. Then
\begin{equation}
	v_{1,i_1} \cdots v_{D, i_D} t_{i_1 \cdots i_D}(\mathbf{u}) =\mathrm{det}~ \left( \begin{array}{c}
		\mathbf{v}_{1}^T\\
		\vdots\\
		\mathbf{v}_{D}^T\\
		\nabla f_1(\mathbf{u})^T\\
		\vdots\\
		\nabla f_E(\mathbf{u})^T\\
	\end{array}\right) = 0,
\end{equation}
where $\nabla$ is the gradient in $\mathbf{u}$ and $^T$ denotes the transpose. Since the $\mathbf{v}_n$ are orthogonal to the other vectors one of the $\nabla f_\alpha(\mathbf{u})$ must be linearly dependent on the rest of them. We immediately obtain that there is at least one additional linear independent zero mode.\\

\noindent \textbf{Self stresses are orthogonal to the critical value set.} The critical set is defined as the set of points $\mathbf{u}_C$ such that $t_{i_1 \cdots i_D}(\mathbf{u}_C) = 0$. The critical value set is the image of the critical set under the map $f_\alpha(\mathbf{u}_C)$. Suppose that $\mathbf{u}_C(s)$ is a one-parameter path of points in a smooth portion of the critical set. Then consider its image $F_\alpha(s)$,
\begin{equation}
	f_\alpha( \mathbf{u}_C(s) ) = F_\alpha(s).
\end{equation}
If the derivative $\partial F_\alpha(s)/\partial s$ is nonzero then it is tangent to the critical value set.  Therefore,
\begin{equation}
	\frac{\partial F_\alpha(s)}{\partial s} = \sum_i \frac{\partial f_\alpha(\mathbf{u}_C(s))}{\partial u_i} \frac{\partial u_{C,i}(s)}{\partial s}.
\end{equation}
If $\sigma_\alpha$ is a self stress then $\sum_\alpha \sigma_\alpha \partial f_\alpha/\partial u_I = 0$. Therefore we obtain
\begin{equation}
	\sum_\alpha \sigma_\alpha \frac{\partial F_\alpha(s)}{\partial s}  = 0.
\end{equation}
Since this is true for any path in the critical value set, it follows that all self stresses are orthogonal to the critical value set.

Though the converse of this is not true -- some vectors normal to the critical value set may not be self stresses -- if the critical value set has codimension one then there can be only one self stress and the normal vector of the critical value set necessarily corresponds to that self stress.\\

\noindent \textbf{Orientation} The tangent form $t_{i_1 \cdots i_D}(\mathbf{u})$ carries additional useful geometrical information about the mechanism at regular (non-critical) configurations. When $D=0$, $t(\mathbf{u})$ is a scalar whose sign was used to compute a topological index in periodic mechanisms \cite{lo2021topology}. Beyond this, it endows the configuration space with a natural orientation in any dimension. At a regular point on the configuration space of a mechanism, $\mathbf{x}$, $t_{i_1 \cdots i_D}(\mathbf{x}) dx^{i_1} \wedge \cdots \wedge dx^{i_D}$ is a differential form which provides a local orientation: for any basis of tangent vectors $\{ \zeta_{1,j}, \cdots, \zeta_{D,j} \}$,
\begin{equation}
    \textrm{sgn}~\sum_{i_1 \cdots i_D} \zeta_{1,i_1} \cdots \zeta_{D,i_D} t_{i_1 \cdots i_D}(\mathbf{u}) = \pm 1
\end{equation}
However, note that this local orientation is only defined up to an overall sign, since we can always take one of the constraint functions to have the opposite sign.

Though we do not make a great deal of use of it in this paper, it is worth noting that if one is able to find two regions in which $t^{i_1 \cdots i_D}(\mathbf{u})$ has opposite signs, there must be a boundary between those regions for which $t^{i_1 \cdots i_D}(\mathbf{u})$ vanishes. That is, in principle we can use the tangent form to verify the existence of critical configurations.

\bibliography{programming}

\end{document}